\documentclass[12pt,letterpaper]{article} \usepackage[latin1] {inputenc} 
\usepackage{amsmath} \usepackage{amsfonts} \usepackage{amssymb} 
\usepackage{amscd}
\begin{document}
\title{Comment on ``A path integral leading to higher-order Lagrangians, arXiv:0708.4351 [hep-th]''}
\author{Ignacio Cortese\footnote{nachoc@nucleares.unam.mx}\,\, and  J. Antonio Garc\'\i a\footnote{ garcia@nucleares.unam.mx}\\
\em Instituto de Ciencias Nucleares, \\
\em Univesidad Nacional Aut\'onoma de M\'exico\\
\em Apartado Postal 70-543, M\'exico D.F., M\'exico}
\maketitle

  \abstract{We show that the ``effective Lagrangian'' constructed in [1] is inconsistent with the exact result for the complete Lagrangian presented in [2]. We trace the origin of the inconsistence to the peculiar way in which the path integral methods are used to eliminate variables that are not auxialiry.}

\section*{}

The new interest in noncommutative physics that comes from a low energy limit of string theory, quantum gravity and noncommutative geometry, has opened old questions in the study of generalized Poisson structures and their possible quantization. Among those questions it remains unclear if a well defined Hamiltonian dynamics based in a noncommutative Poisson structure admits a variational formulation in configuration space. It is well known that even in the simplest case where the Poisson structure has the form
\begin{equation}
\label{sym-xp}
\{q^i,q^j\}=\theta^{ij},\quad \{p_i,p_j\}=0,\quad \{q^i,p_j\}=\delta^i_j,
\end{equation}
with $\theta$ a constant matrix,  the associated problem in configuration space is not variationally admissible. That means that it is not possible to construct a standard Lagrangian in configuration space that reproduce the dynamics of the corresponding equations of motion. This question is not only interesting from the point of view of  the inverse problem of the calculus of variations but it is also interesting for  physical reasons  because we are ultimately interested in the description of the dynamical properties of the system in the noncommutative configuration space.

In this context we want to comment on the recent work \cite{acatri} where the author claims that an ``effective'' Lagrangian in configuration space for a Hamiltonian system defined by (\ref{sym-xp}) and $H=T+V$ can be constructed. It turns out that the Lagrangian is linear on the accelerations and consequently the equations of motion are of third order in derivatives with respect to time. Unfortunately these equations of motion are not equivalent to the original set of equations of motion. It is not clear from \cite{acatri} in what sense the dynamical description of the original problem is effective. At least, we can say that it is not a consistent truncation up to terms linear in $\theta$ of the full variational description (see below).  

The author of \cite{acatri} is apparently unaware of the recent result obtained in \cite{CG} where a complete solution of the inverse problem of the calculus of variations for the original equations of motion (see \ref{nc-f}) was found.

Let us recall the variational construction as presented in \cite{CG}. 
The Hamiltonian equations are
 \begin{subequations}
\label{h-eq}
\begin{align}
\dot q^i + \theta^{ij}\dot{p_j}- p_i=0,\label{he-a}\\
-\dot{p_i} -\frac{\partial V}{\partial q^i}=0.\label{he-b}
\end{align}
\end{subequations}
 Playing with the equations of motion alone it is easy to show that the equations of motion in configuration space are
  \begin{equation}
\label{nc-f}
{\ddot q}^i+\frac{\partial V}{\partial q^i}-\theta^{ij}\frac{d}{dt}\frac{\partial V}{\partial q^j}=0.
\end{equation}
It is clear from here that the momenta are not auxiliary variables as in the standard formulation of Hamiltonian dynamical systems.

  A Lagrangian whose equations of motion are equivalent to the dynamics described by (\ref{nc-f}) is \footnote{A generalization to field theory of this idea has been presented in \cite{lyan}.} (see ref. \cite{CG} for details)
 \begin{equation}\label{nc-L}
L(\ddot q,\dot q,q)=L_0-\frac12\theta^{ij}\dot q^i \dot V^j+\frac12\theta^{ij}    
\ddot q^j(\dot q^i-V^i)+\frac12\theta^{ij}V^i\dot V^j-\frac12 V^kV^k,
\end{equation}
where $L_0$ is the standard commutative Lagrangian $L_0=T-V$ and
$$V^i\equiv\theta^{ij}V_j=\theta^{ij}\frac{\partial V}{\partial q^j}.$$

 It is easy to verify that the equations of motion associated with the Lagrangian (\ref{nc-L}) are
 \begin{equation}\label{tode2}
C_{ij}(\ddot q^j-F^j(q,\dot q))^{\cdot}-A_{ij}(\ddot q^j-F^j(q,\dot q))=0,
\end{equation}
where
\begin{subequations}
\begin{align}
&F^i=-\frac{\partial V}{\partial q^i}+\theta^{ij}\frac{d}{dt}\frac{\partial V}{\partial q^j},\\
&A_{ij}=-\delta_{ij}+\theta^{jk}V^k_i,\\ 
&C_{ij}=\theta^{ij}.
\end{align}
\end{subequations}
 We must keep in mind that the extended derivative along the solutions of the system (\ref{tode2})  is
 \begin{equation}
 \label{flux}
 \frac{D}{Dt}={\dot F}^i\frac{\partial}{\partial{\ddot q}^i} +
{F}^i\frac{\partial}{\partial{\dot q}^i}+
{\dot q}^i\frac{\partial}{\partial q^i}+
\frac{\partial}{\partial t}.
\end{equation}

 The equations (\ref{tode2}) are of third order but the crux of the argument is that the variational description given in terms of (\ref{nc-L}) depends in a fundamental way of the fact that we are restricted to the definition of the dynamical flux given by (\ref{flux}). Without this basic condition the solution space of the equations of motion that come from the Lagrangian (\ref{nc-L}) is bigger than the solution space of the original equations of motion.

To end our comment we observe that the peculiar way to use the path integral in \cite{acatri} to eliminate the momenta from the description of the Hamiltonian dynamics has at least two drawbacks, 
a) The transformation of variables (9) of \cite{acatri} is non canonical with respect to the Poisson bracket (\ref{sym-xp}). But the author claims that the path integral measure is invariant! b) The path integral needs a proper and very careful adjustment of the boundary data because it is in the boundaries that reside the information of the kernel to be calculated. It is obvious that when the configuration variables does not commute the operation to integrate out the momenta is not allowed.
Trying to justify the construct of \cite{acatri} from a proposal of ``A path integral...'' with quite non standard rules seems to us completely unecessary.

\end{document}